\documentclass[preprint,prd,aps,pdftex,showpacs,preprintnumbers,amsmath,amssymb]{revtex4-1}
\usepackage{graphics,epsfig,subfigure}
\usepackage{diagbox}
\usepackage[usenames]{color}
\usepackage{color}
\usepackage{graphicx}
\usepackage{amsfonts}
\usepackage{indentfirst}
\usepackage{booktabs}
\usepackage{hyperref}
\hypersetup{hypertex=ture,backref=true,colorlinks=ture,linkcolor=blue,anchorcolor=blue,citecolor=blue}
\usepackage{float}
\usepackage{multirow}

\begin{document}

\title{Compact Stars Sourced by Dark Matter Halos and Their Frozen States}

\author{Yuan Yue$^{1}$}
\author{Yong-Qiang Wang$^{2}$}\email{yqwang@lzu.edu.cn, corresponding author}
\affiliation{$^{1}$College of Mathematics and Computer Science, Northwest Minzu University, Lanzhou, 730030, China\\
$^{2}$School of Physical Science and Technology, Lanzhou University, Lanzhou 730000, China}

\begin{abstract}
Inspired by regular black holes (RBHs) sourced by dark matter  halos, we generalize the anisotropic energy-momentum tensor by relaxing the $P_r = -\rho$  condition between radial pressure and density. We demonstrate that while RBHs are a unique special case, a broader class of relations yields horizonless compact stars. Under specific parameter limits, these objects approach a  ``frozen state," mimicking black hole features without an event horizon. These compact star solutions could satisfy weak energy conditions and provide a robust mechanism for dark matter-sourced black hole mimickers.
\end{abstract}

\maketitle

\section{Introduction}
The nature of singularities in general relativity remains one of the most profound enigmas in modern theoretical physics. The singularity theorems of Penrose and Hawking establish that the gravitational collapse of normal matter inevitably leads to spacetime singularities [1,2]. It is widely anticipated that these singularities are artifacts of the classical theory and will be resolved by quantum-gravity effects or through appropriate matter sources that regularize the geometry.

The construction of singularity-free black holes has attracted sustained interest for decades. The first such example was proposed by Bardeen \cite{Bardeen:1968} as a static, spherically symmetric spacetime that replaces the central singularity with a de Sitter-like core while preserving an event horizon and asymptotic flatness. This geometry was subsequently shown to arise as an exact solution of Einstein equations coupled to a nonlinear magnetic monopole in a specific form of nonlinear electrodynamics \cite{Ayon-Beato:1998hmi, Ayon-Beato:2000mjt}.
In recent years, regular black holes arising from purely geometric constructions—independent of matter sources—have been discovered. Driven by advances in higher-derivative gravity, a new class of such black holes has emerged within the framework of infinite-order quasi-topological gravity extensions \cite{Bueno:2024dgm,Bueno:2025zaj}.

More recently, a novel family of regular black hole solutions has been derived by sourcing the geometry with dark matter halos described by realistic galactic density profiles, notably the Einasto and Dehnen models \cite{Konoplya:2025}. This framework naturally connects small-scale black hole physics with large-scale galactic dynamics and offers a phenomenological route to embedding singularity-free horizons in DM-dominated environments. By imposing $  P_r = -\rho  $ between radial pressure and density, these models resolve the central singularity while preserving asymptotic flatness and consistency with galactic boundary conditions. This discovery has led to extensive investigations into their observational signatures, including quasinormal modes (QNMs), grey-body factors, and absorption cross-sections across various fields \cite{Lutfuoglu:2025grav, Bolokhov:2025qnm, Saka:2025, Malik:2025}. 
However, the relation $  P_r = -\rho  $ represents a particular choice within the  parameter space of anisotropic fluid dynamics. In self-gravitating systems, there is no fundamental justification for requiring the radial pressure to precisely cancel the energy density everywhere. Recent analyses \cite{
Bolokhov:2025crit} provided a critical review of various black hole metrics embedded in dark-matter halos, pointing out that many existing solutions fail to consistently satisfy the Einstein equations for their intended matter distributions.

In this work, we systematically relax the constraint between radial pressure and density, treating $  P_r = -\rho  $ as merely a special case. This generalization allows us to obtain fully self-consistent solutions to the Einstein equations in a broader anisotropic setting.
By exploring more general relations between density and pressure components, we gain a deeper insight into the solution space and uncover a variety of horizonless compact objects. In particular,  We show that the $  P_r = -\rho  $ case, which produces regular black hole geometries \cite{Konoplya:2025}, emerges as a special case of our analyses. For the more general relations, the resulting  spacetimes are horizonless, describing a novel class of compact stars sourced by dark matter.
Moreover, under suitable parameter regimes, these configurations can approach a ``frozen state" in which the metric function $  g_{tt} $ approaches zero arbitrarily closely at a finite radius without the formation of a conventional event horizon, thereby closely mimicking black hole properties from the perspective of distant observers,  which also were found in both boson stars \cite{Wang:2023tdz,Yue:2023sep,Chicaiza-Medina:2025wul} and neutron stars \cite{Tan:2025jcg} within the Bardeen and Hayward models. Such frozen objects are of particular interest as potential black hole mimickers.

The organization of the paper is as follows. In Sec. II, we introduce the generalized energy-momentum tensor and the corresponding Einstein field equations. Section III presents numerical and analytical solutions for the compact star configurations under various DM density profiles. In Sec. IV, We investigate the stability of the spacetime background under axial perturbations. Finally, Sec. V summarizes our results.

\section{Model Setup}
In this section, we provide an introduction to the theoretical framework. The gravitational field is governed by the Einstein field equations with mixed indices, which relate the spacetime geometry to the energy-momentum tensor $T^{\mu}{}_{\nu}$ of the dark matter fluid:
\begin{equation}\label{eom1}
    G^{\mu}{}_{\nu} = R^{\mu}{}_{\nu} - \frac{1}{2} \delta^{\mu}{}_{\nu} R = 8\pi T^{\mu}{}_{\nu}.
\end{equation}
For a static and spherically symmetric configuration with anisotropic pressure, the mixed-index energy-momentum tensor is given by the diagonal form:
\begin{equation}
    T^{\mu}{}_{\nu} = \text{diag}(-\rho, P_r, P_\perp, P_\perp),
    \label{eq:EnergyMomentum}
\end{equation}
where $\rho(r)$ is the energy density, $P_r(r)$ is the radial pressure, and $P_\perp(r)$ is the tangential pressure. 
We contemplate a generic static spherically symmetric solution and employ the following ansatz:
\begin{equation}\label{equ10}
	ds^2 = -N(r)\sigma^2(r)dt^2 + \frac{dr^2}{N(r)} + r^2\left(d\theta^2 + \sin^2\theta d\varphi^2\right).
\end{equation}
Herein, the functions $N(r)$ and $\sigma(r)$  depend exclusively  on the radial variable $r$. 
By substituting the subsequent ansatze for the metric and energy-momentum tensor into the Einstein field equations (\ref{eom1}), we obtain the following set of gravitational field equations for the anisotropic dark matter:
\begin{align}
  G^t{}_t &=   -8\pi \rho = \frac{1}{r^2} \left( r N' + N - 1 \right), \label{eq:field1} \\
  G^r{}_r &=   8\pi P_r = \frac{1}{r^2} \left[ N \left( \frac{2r\sigma'}{\sigma} + 1 \right) + r N' - 1 \right], \label{eq:field2} \\
  G^\theta{}_\theta = G^\phi{}_\phi &=   8\pi P_\perp = \frac{N''}{2} + \frac{N'}{r} + \frac{3 N' \sigma'}{2 \sigma} + \frac{N \sigma''}{\sigma} + \frac{N \sigma'}{r \sigma}, \label{eq:field3}
\end{align}
where the prime denotes the derivative with respect to the radial coordinate $r$. 
Furthermore, the energy-momentum tensor of the dark matter halo must satisfy the conservation law $\nabla_\mu T^\mu{}_\nu = 0$. For a static and spherically symmetric metric, the only non-trivial component of the conservation equation is the radial one ($\nu = r$), which yields the generalized Tolman-Oppenheimer-Volkoff (TOV) equation for an anisotropic fluid:
\begin{equation}
    P_r' + \frac{1}{2} \left( \frac{N'}{N} + \frac{2\sigma'}{\sigma} \right) (\rho + P_r) + \frac{2}{r} (P_r - P_\perp) = 0. \label{eq:TOV}
\end{equation}
Eqs. \eqref{eq:field1}, \eqref{eq:field2}, \eqref{eq:field3}, and \eqref{eq:TOV} constitute a set of coupled equations. Since the ${}^\theta{}_\theta$ component of the Einstein equations is not independent and can be derived from the aforementioned equations, we shall solve the following three equations:
\begin{align}
   N' &= \frac{1-N - 8\pi r^2 \rho}{r}, \label{eq:N_prime} \\
    \sigma' &= \frac{4\pi r \sigma (\rho + P_r)}{N}, \label{eq:sigma_prime} \\
    P_\perp &= P_r + \frac{r}{2} P_r' + \frac{r(\rho + P_r)}{4} \left( \frac{N'}{N} + \frac{2\sigma'}{\sigma} \right). \label{eq:sigma_double_prime}
\end{align}

In order to solve the above set of ordinary differential equations, it is necessary to first specify the explicit functional forms of $\rho$ and $P_r$. Notably, under the special condition $\rho + P_r = 0$, the equations could reduce to the cases  in \cite{Konoplya:2025}. Furthermore, it is important to specify suitable boundary conditions for each unknown function. Given the asymptotically flat characteristics of the solutions, the metric functions $N(r)$ and $\sigma(r)$ are required to satisfy the boundary conditions:
\begin{equation}\label{equ19}
N(0) = 1,\qquad N(\infty) = 1-\frac{2 M}{r}, \qquad \sigma(\infty) = 1.
\end{equation}
Furthermore, specific boundary conditions are mandated for the tangential
pressure
\begin{equation}\label{equ20}
P_\perp(\infty) = 0,
\end{equation}

In the subsequent calculations, we also adopt an Einasto-type density profile \cite{Einasto1965,EinastoHaud1989,Retana2012,Konoplya:2025}:
\begin{equation}
 \rho(r) = \rho_{0} \exp \left[ -\left( \frac{r}{h} \right)^{1/n} \right], \quad n > 0.
\end{equation}
For the radial pressure $P_r(r)$, subject to the boundary condition $P_r(\infty) = 0$, we consider the following two distinct scenarios:
\begin{itemize}
    \item \textbf{Case I:} A barotropic-like relation where the pressure is proportional to the density,
    \begin{equation}\label{pr1}
     P_r(r) = a_0 \,\rho(r), \quad a_0 > -1.
    \end{equation}
    \item \textbf{Case II:} A generalized profile characterized by an exponential-power-law decay,
    \begin{equation}\label{pr2}
     P_r(r) = P_{0} \exp \left[ -(a r)^{m} \right] \left( 1 + \left(\frac{r}{b}\right)^{k+1} \right)^{-\gamma},
    \end{equation}
    where the parameters $\{P_0, a, m, b, k, \gamma\}$ are constants that determine the pressure distribution.
\end{itemize}

\section{Numerical results}\label{sec3}
In this section, we numerically integrate the coupled system of equations (\ref{eq:N_prime}), (\ref{eq:sigma_prime}), and (\ref{eq:sigma_double_prime}) subject to the boundary conditions defined in Eqs. (\ref{equ19}) and (\ref{equ20}). To facilitate the numerical treatment, we map the infinite radial domain $r \in [0, \infty)$ onto a finite interval $x \in [0, 1]$ by introducing the compactified coordinate $x = \frac{r}{1+r}$.

\subsection{Case I}
\begin{figure}
  \includegraphics[width=9cm]{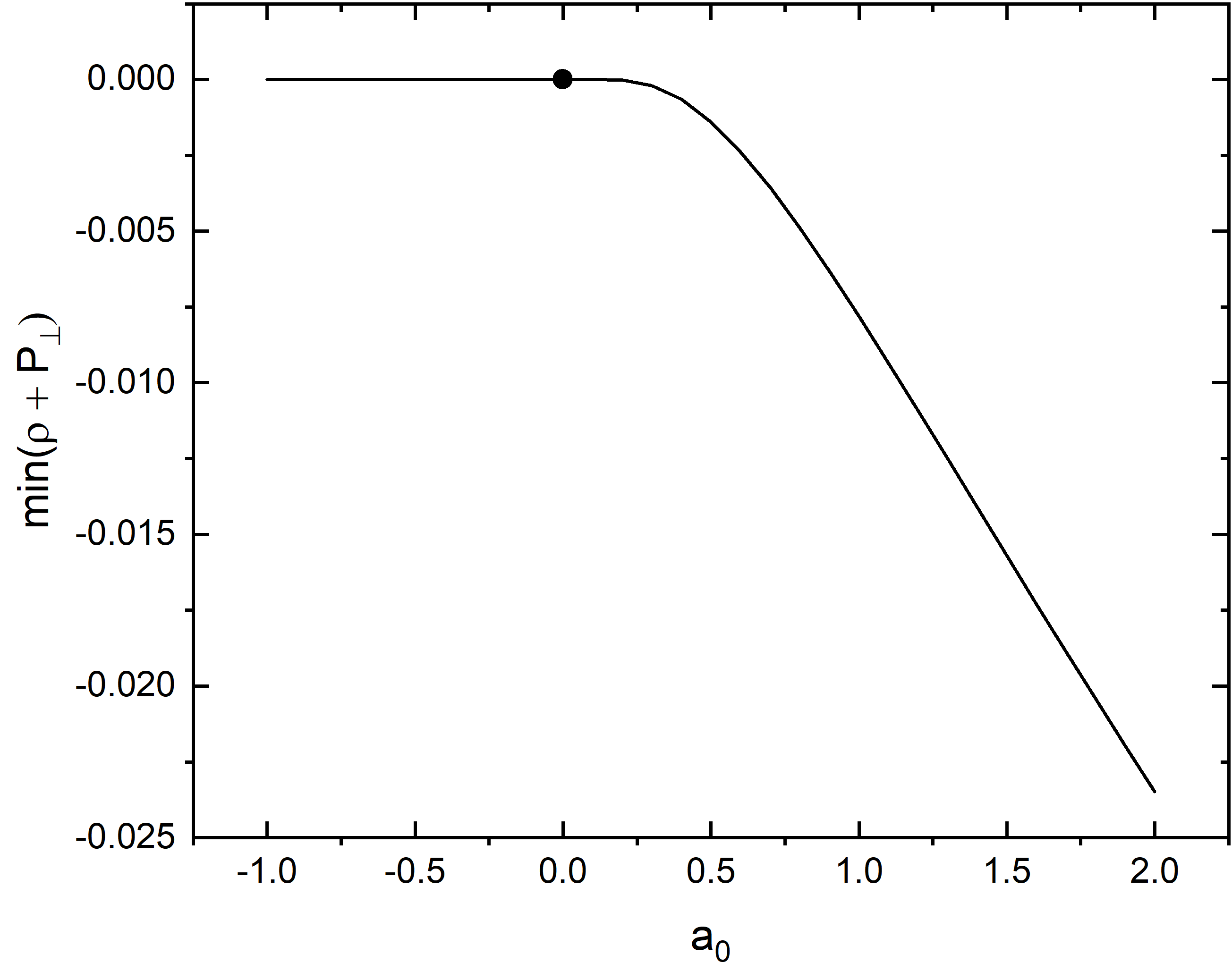}
  \caption{The relationship between the minimum of $\rho +P_\perp$ and  $a_0$ with the fixed parameters $\rho_0=0.5$,\; $n=0.5$ and $h = 0.5 $. }\label{p-a}
\end{figure}

In this case, we adopt the radial pressure $P_r(r)$ as defined in Eq. (\ref{pr1}). By solving the coupled gravitational field equations, we first show the relationship between the minimum of $\rho +P_\perp$ and the parameter $a_0$, as illustrated in Fig. \ref{p-a}, to examine the Weak energy condition (WEC). Our numerical results indicate that the weak energy condition is strictly satisfied when the parameter $a_0$ lies in the range $-1 < a_0 \leq 0$. Conversely, for values of $a_0 > 0$, the energy condition is violated. Based on this observation, and to ensure the physical viability of our configurations, we restrict our subsequent calculations and analysis to the regime where $-1 < a_0 \leq 0$.

\begin{figure}
  \includegraphics[width=8.2cm]{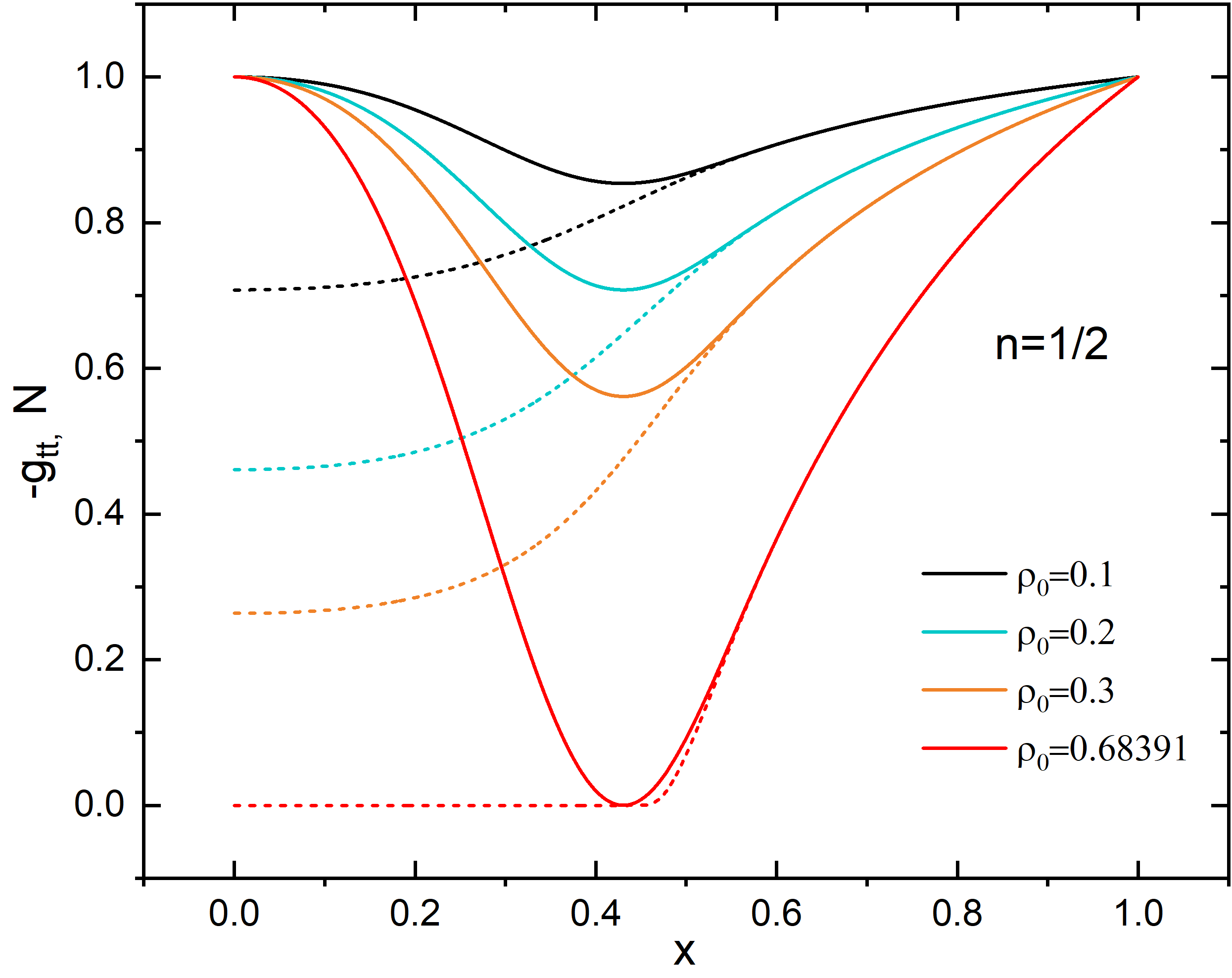}
 \includegraphics[width=8.1cm]{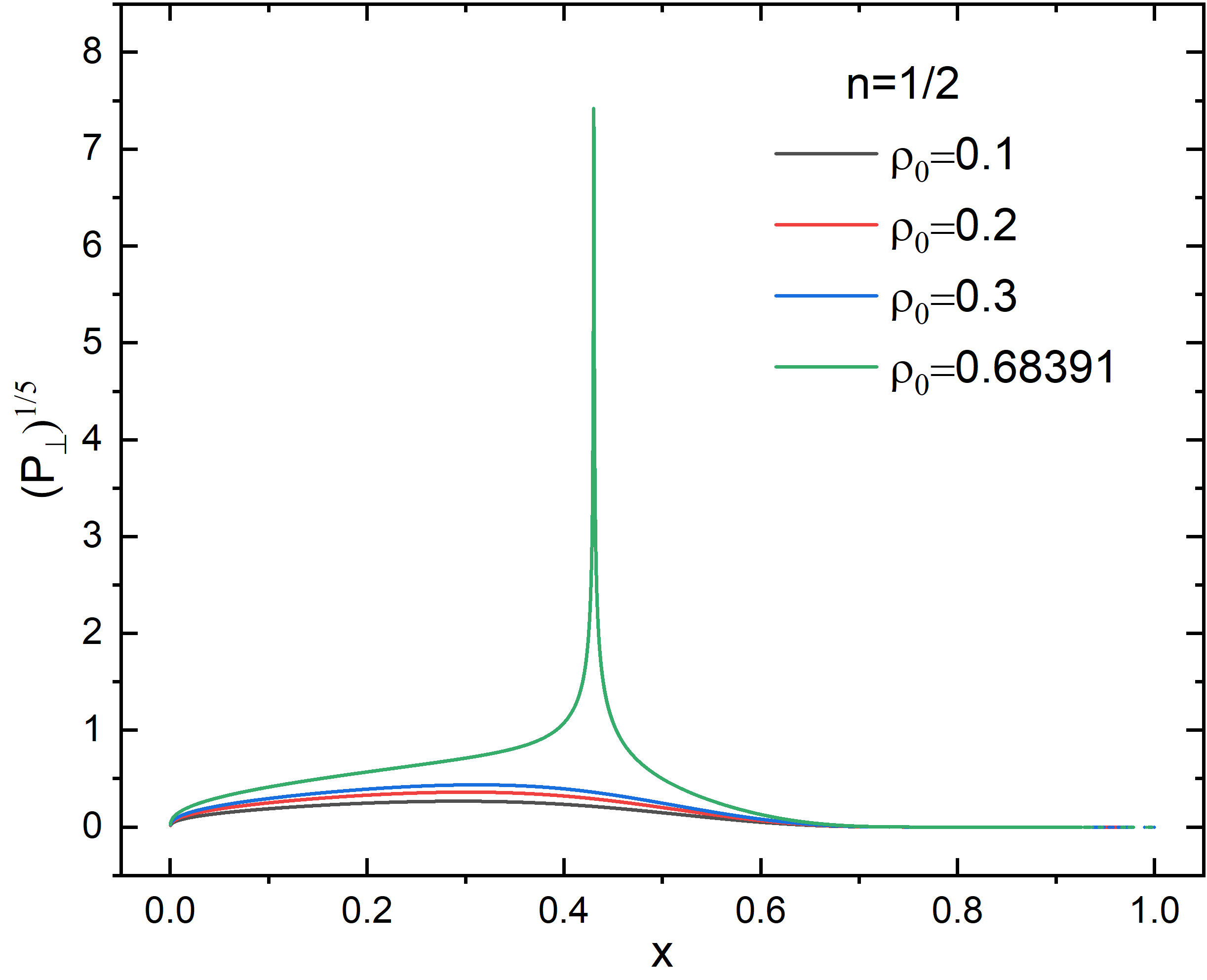}
  \begin{center}
      \includegraphics[width=8.1cm]{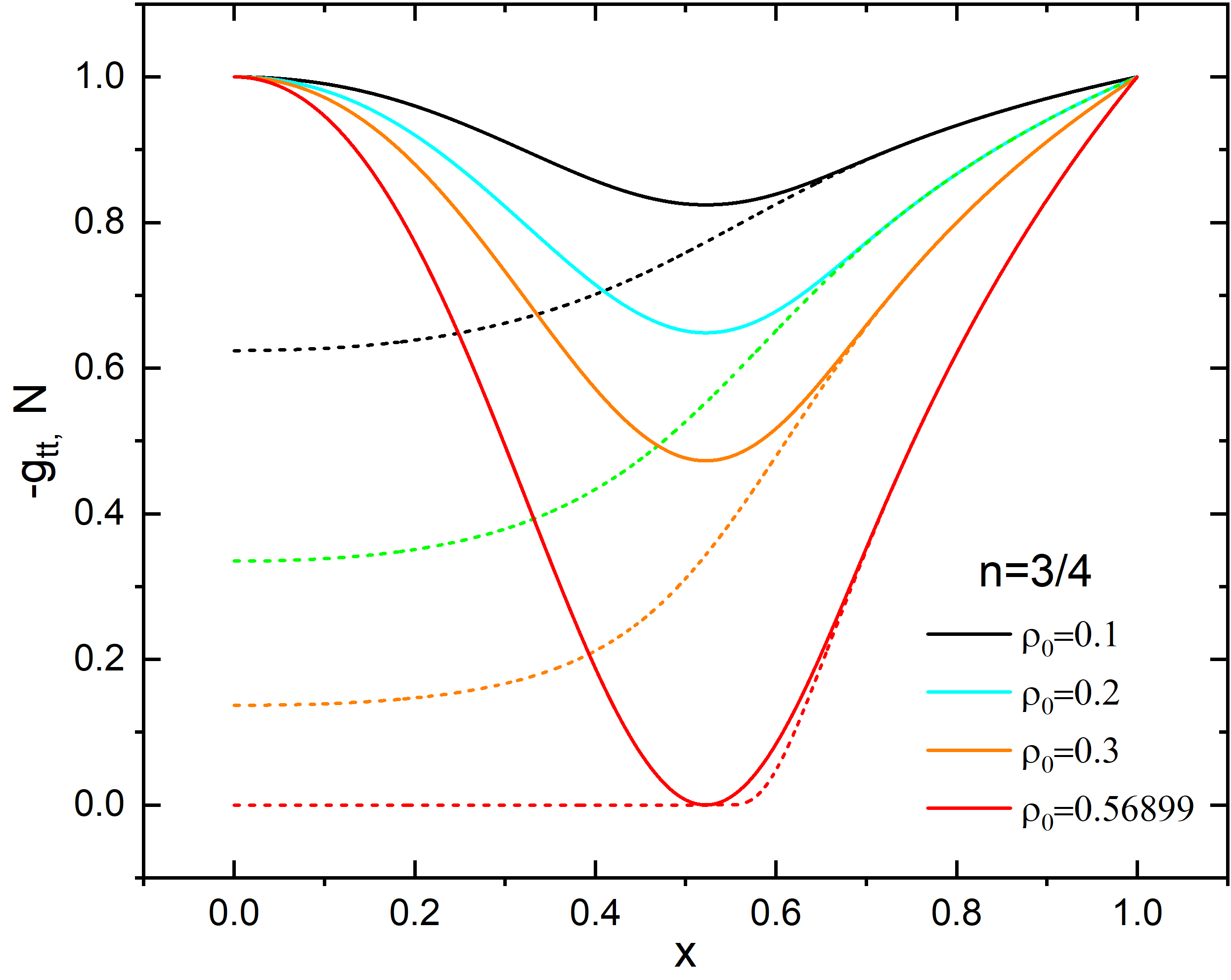}    \includegraphics[width=8.1cm]{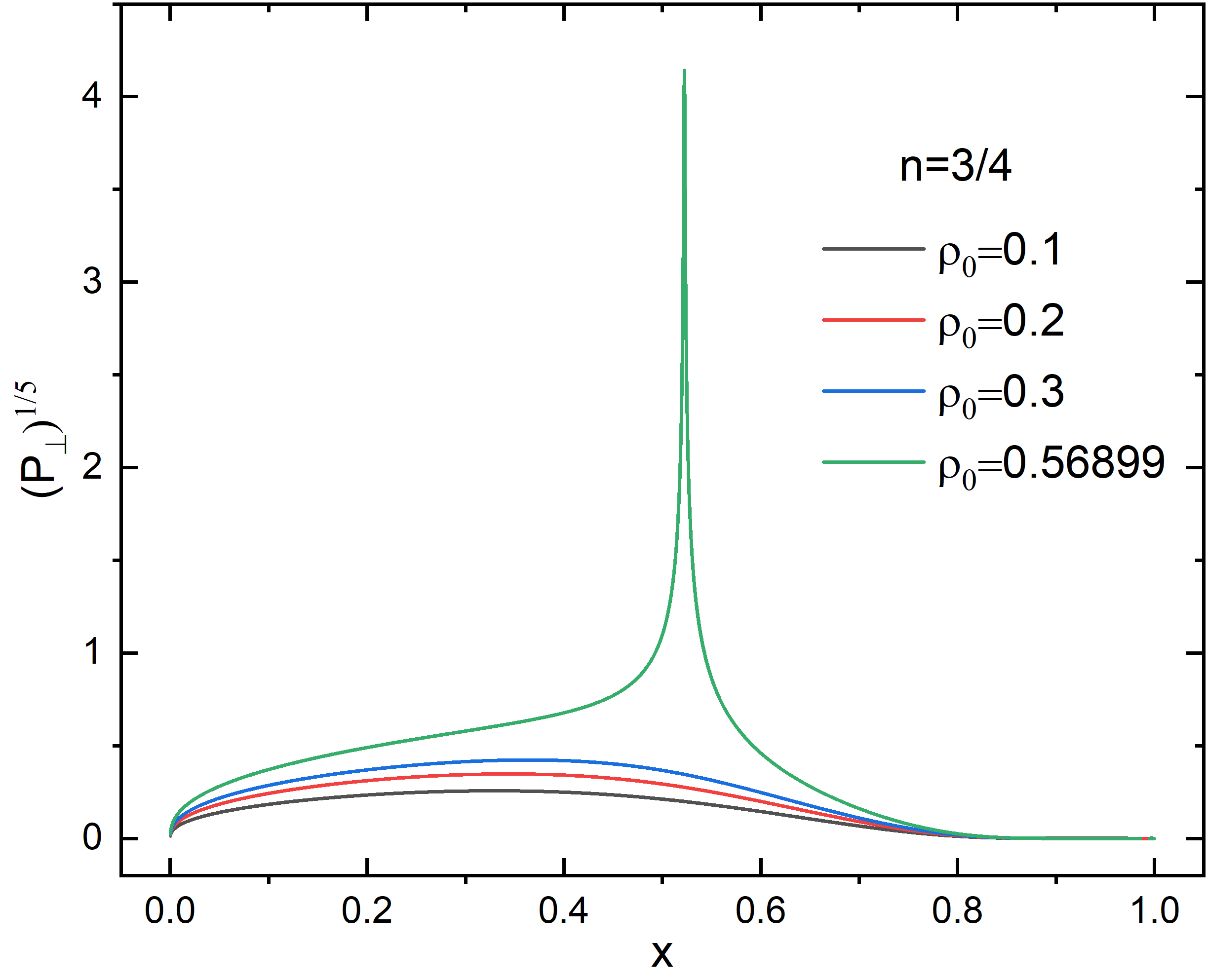}   \includegraphics[width=8.1cm]{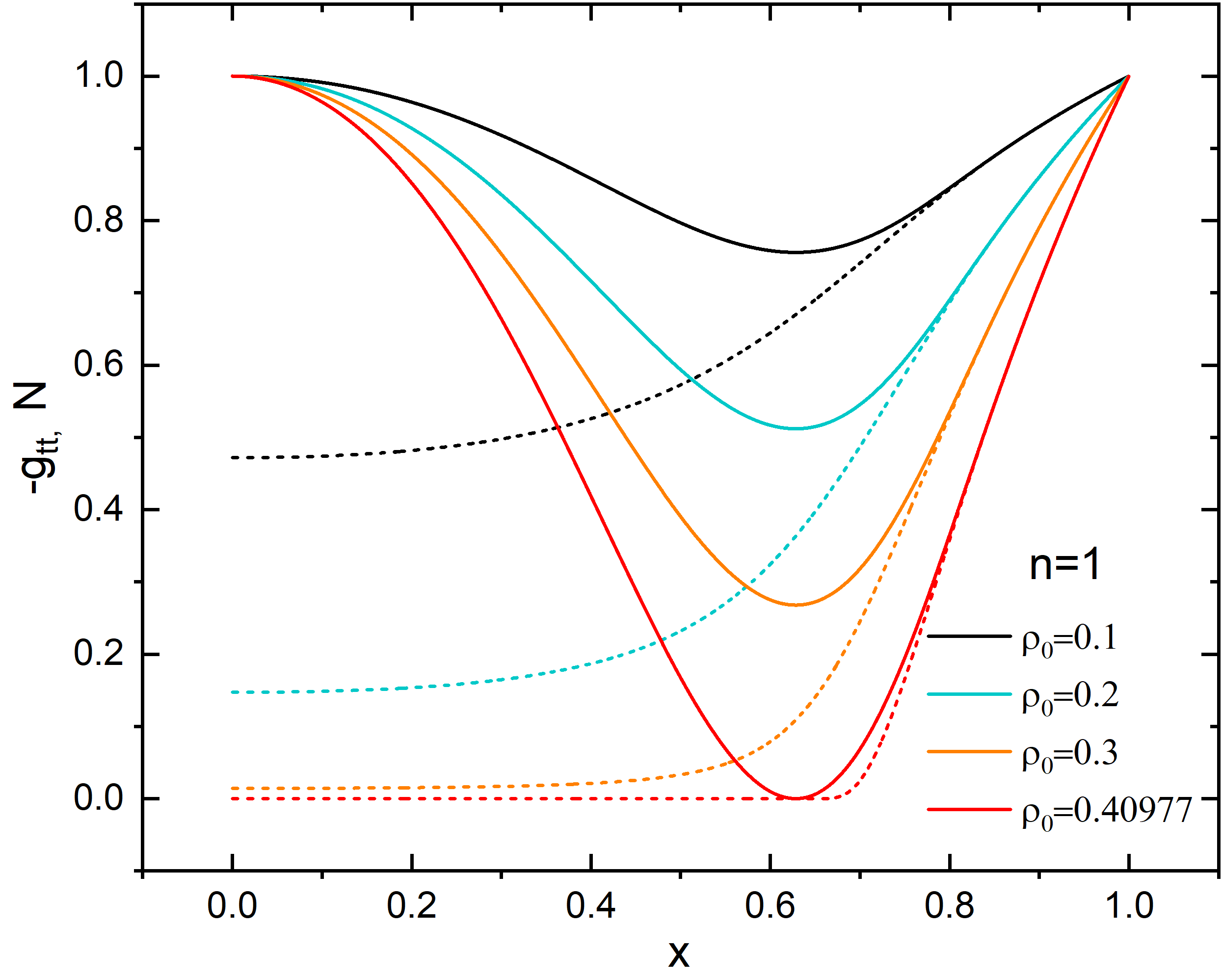}
      \includegraphics[width=8.1cm]{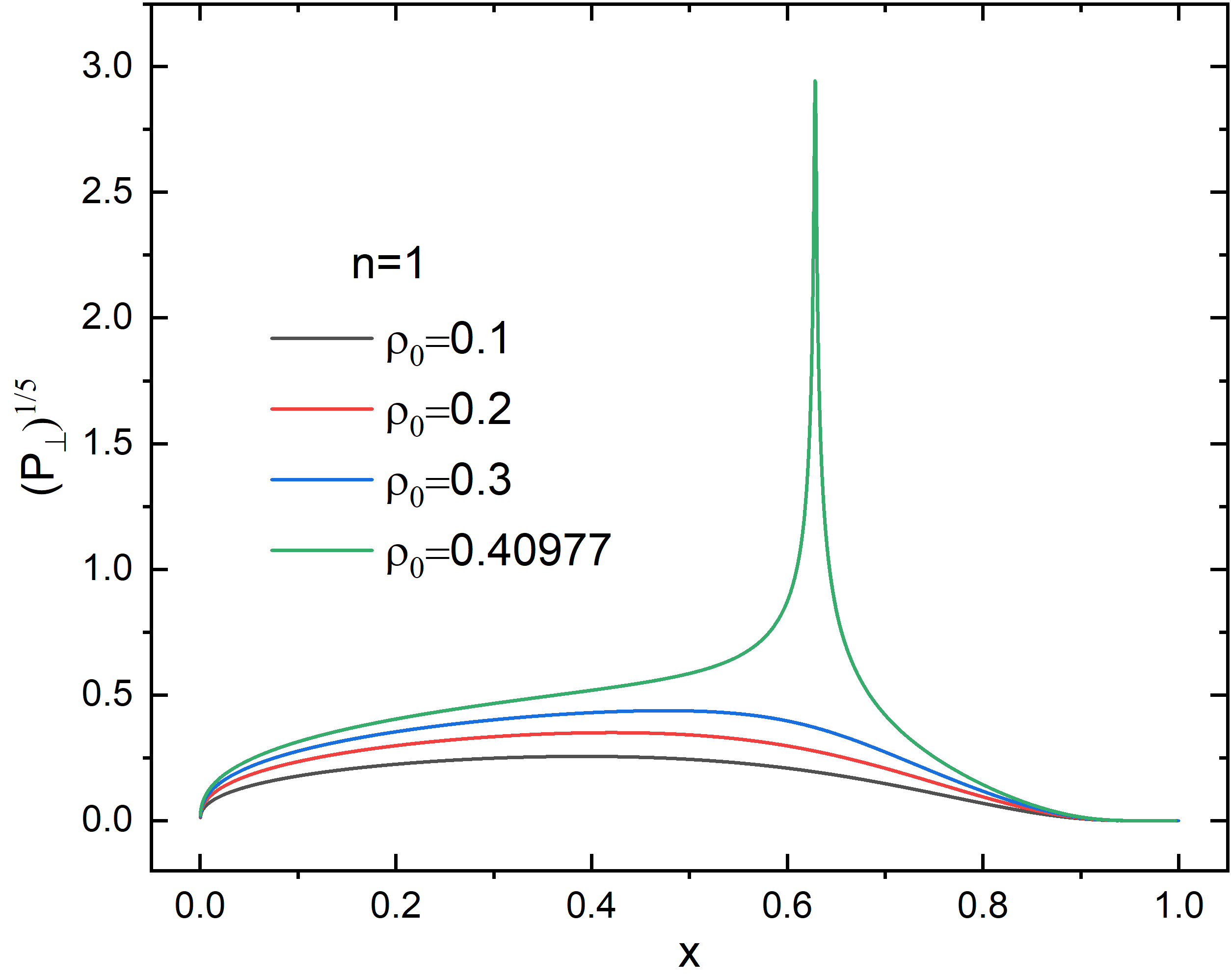}
  \end{center}
  \caption{The spatial distributions of functions $-g_{tt}$, $N$, and $P_\perp$ for different values of $\rho_0$ with $a_0=0$ and $h=0.5$.
  }\label{phase}
\end{figure}

In Fig. \ref{phase},  we present the spatial distributions of functions $-g_{tt}$, $N$, and $P_\perp$ for different values of $\rho_0$ with the fixed parameters $a_0=0$ and $h=0.5$. 
The left column of Fig. \ref{phase} displays the spatial distributions of the metric functions $-g_{tt}$ (dashed lines) and $N$ (solid lines) for different values of $n = 1/2, \; 3/4,\; 1$. It is observed that as $\rho_0$ increases, the minimum value of $-g_{tt}$ decreases significantly. There exists a maximum threshold for $\rho_0$ at which the minimum of $-g_{tt}$ asymptotically approaches zero. We define the corresponding radial coordinate as the critical radius, and the resulting configuration is referred to as a \textit{frozen star}. Specifically, for $n = 1/2,\; 3/4,\;$ and $1$, the critical densities are found to be $\rho_0 = 0.68391, 0.56899,$ and $0.40977$, respectively. These values coincide with the minimum density requirements for the existence of regular extremal black holes.
In the right column, we illustrate the spatial distribution of the tangential pressure $P_\perp$. As the system approaches the frozen state, $P_\perp$ exhibits a prominent peak near the critical radius. Notably, this peak grows indefinitely as $\rho_0$ nears the critical limit, signaling a divergent behavior in the tangential pressure as the frozen star configuration is realized.
\begin{figure}[]
  \begin{center}
  \includegraphics[width=8.1cm]{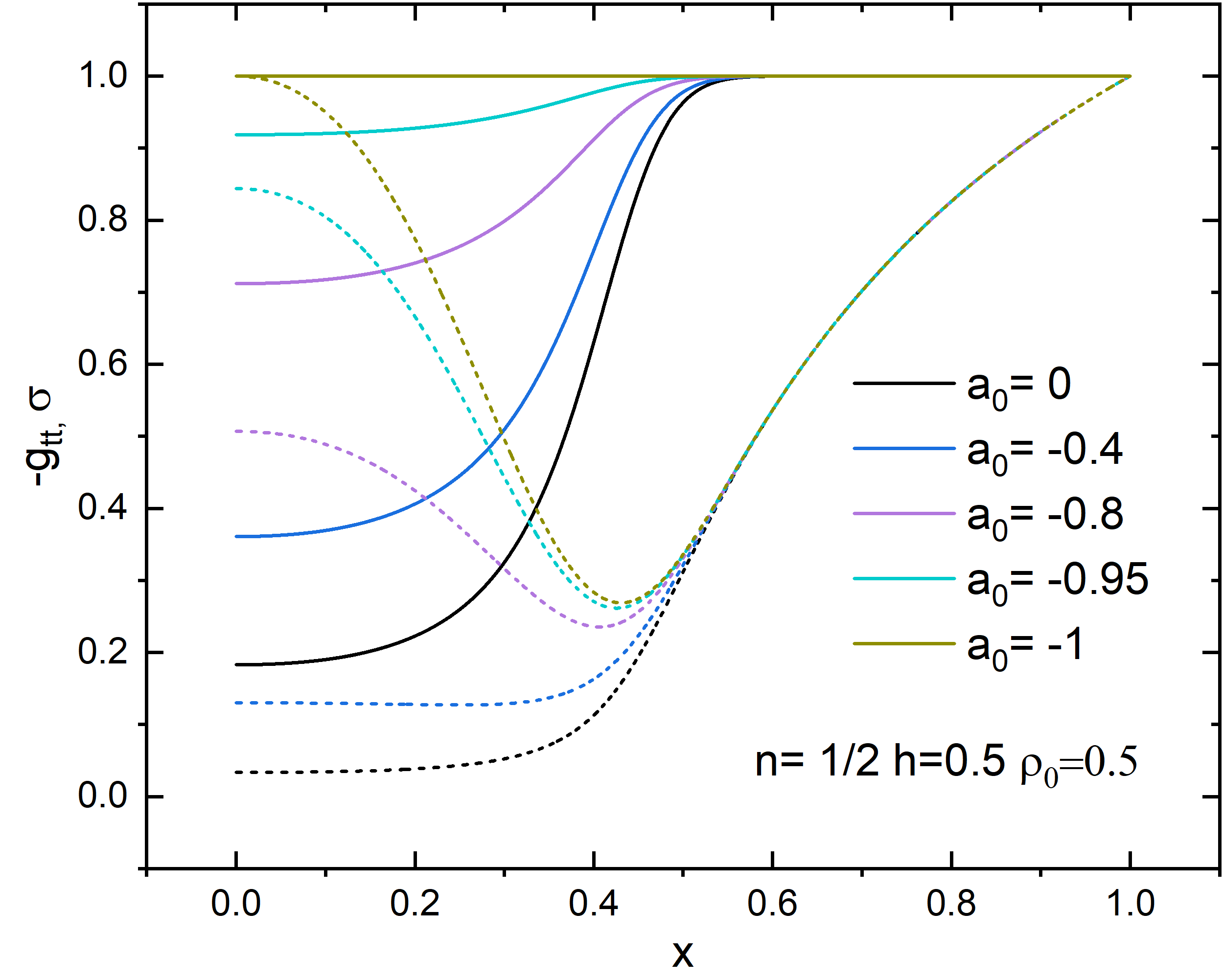}
   \includegraphics[width=8.1cm]{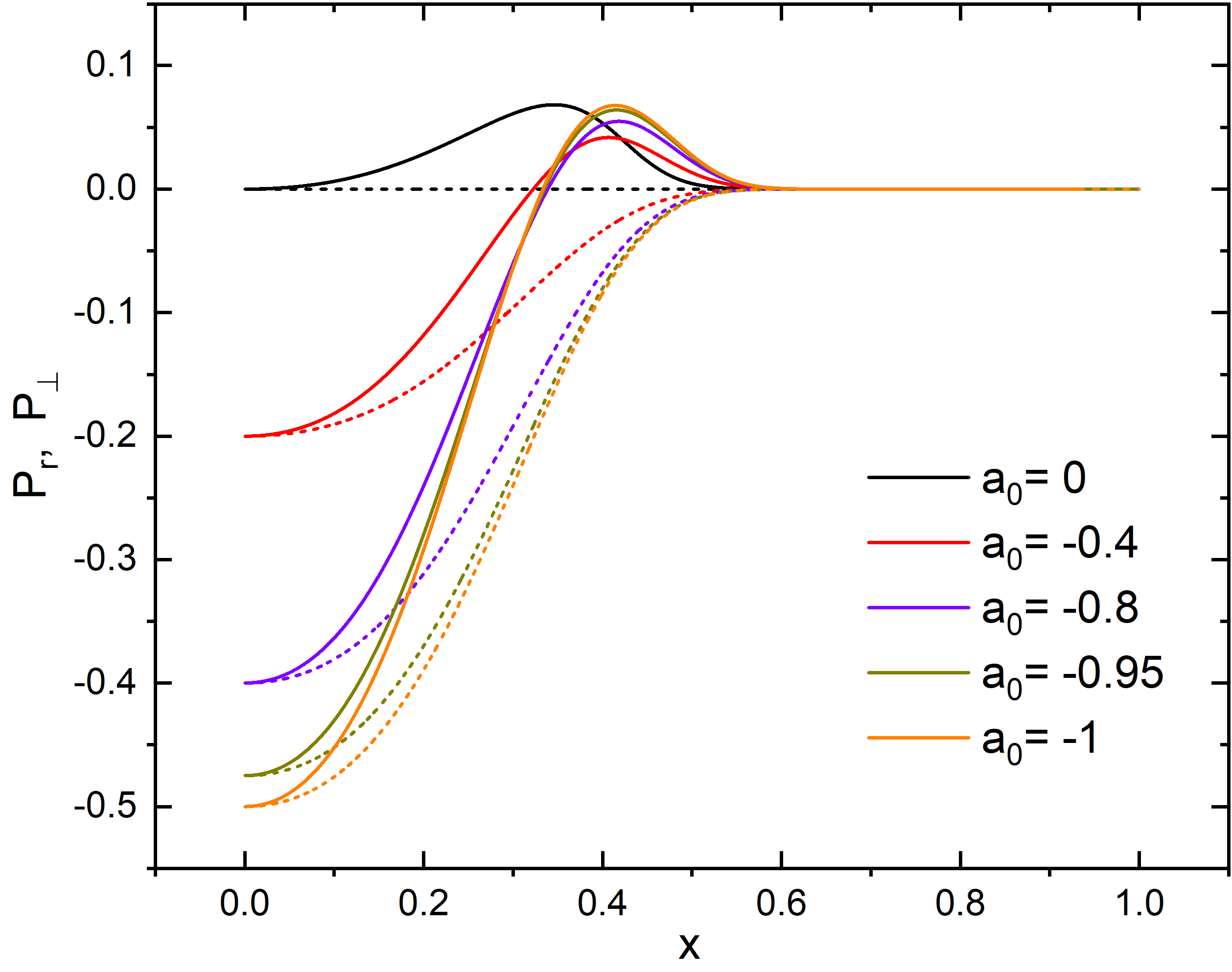}
  \end{center}
  \caption{Spatial distributions of the metric functions $-g_{tt}$ (dashed lines), $\sigma$ (solid lines), the radial pressure $P_r$ (dashed lines)  and tangential pressure $P_\perp$  (solid lines) under varying $a_0$. 
  }\label{gttsigpru1}
\end{figure}

To further explore the role of the radial pressure $P_r(r)$, we examine the effects of varying $a_0$ while keeping other parameters fixed. The left panel of Fig.~\ref{gttsigpru1} displays the spatial distributions of the metric functions $-g_{tt}$ (dashed lines) and $\sigma$ (solid lines)  under varying $a_0$. When $a_0 = -1$, corresponding to the condition $\rho + P_r = 0$, the solutions reduce to the regular black hole cases in ~\cite{Konoplya:2025}. As $a_0$ increases, the minimum value of $-g_{tt}$ and $\sigma$ decreases significantly,  indicating a transition toward more compact configurations.
The right panel shows the radial pressure $P_r$ and tangential pressure $P_\perp$. As $a_0$ increases, the amplitude of $P_r$ decreases progressively until it vanishes for $a_0 = 0$.

\begin{figure}[]
  \begin{center}
  \includegraphics[width=8.1cm]{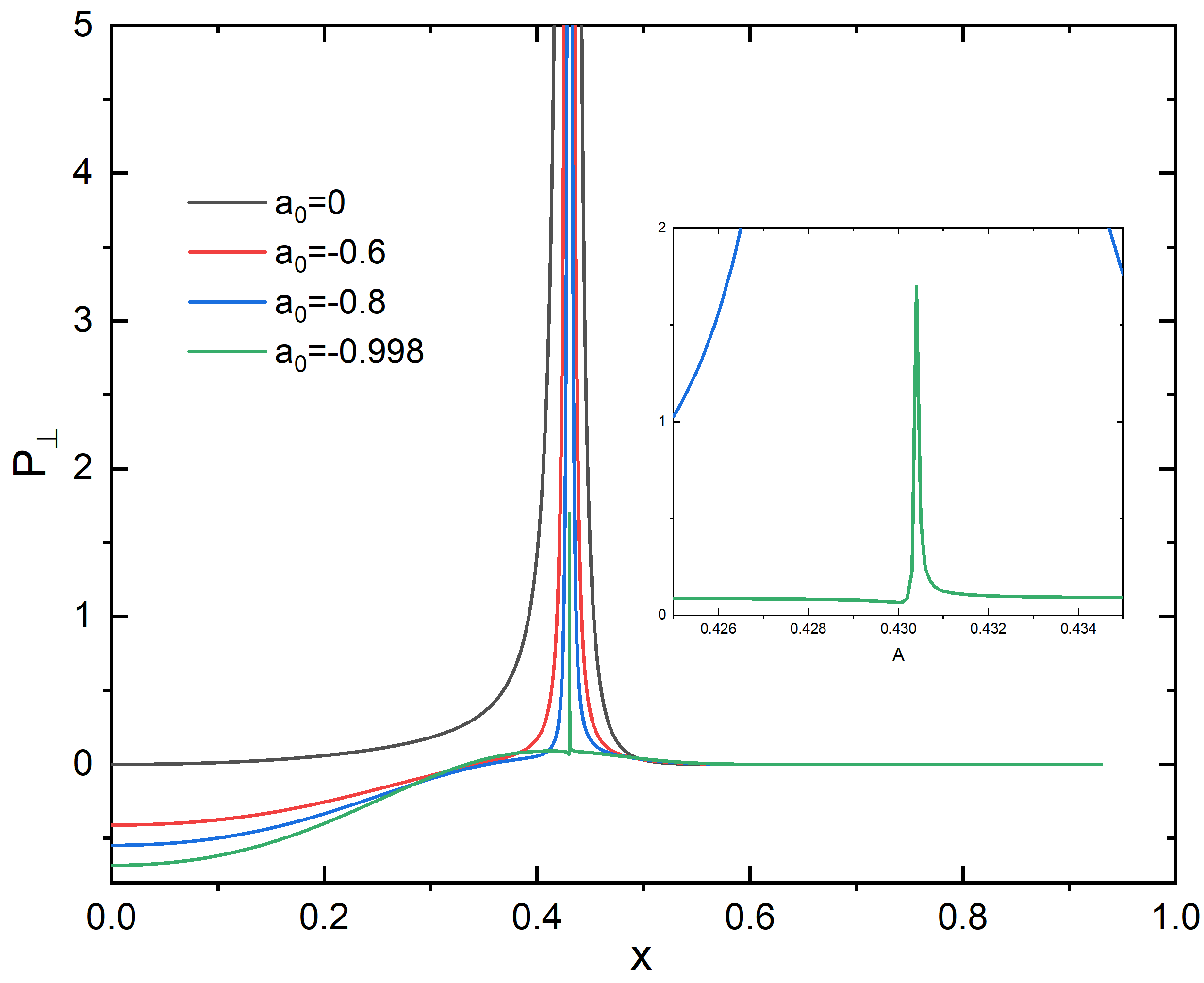}
   \includegraphics[width=8.1cm]{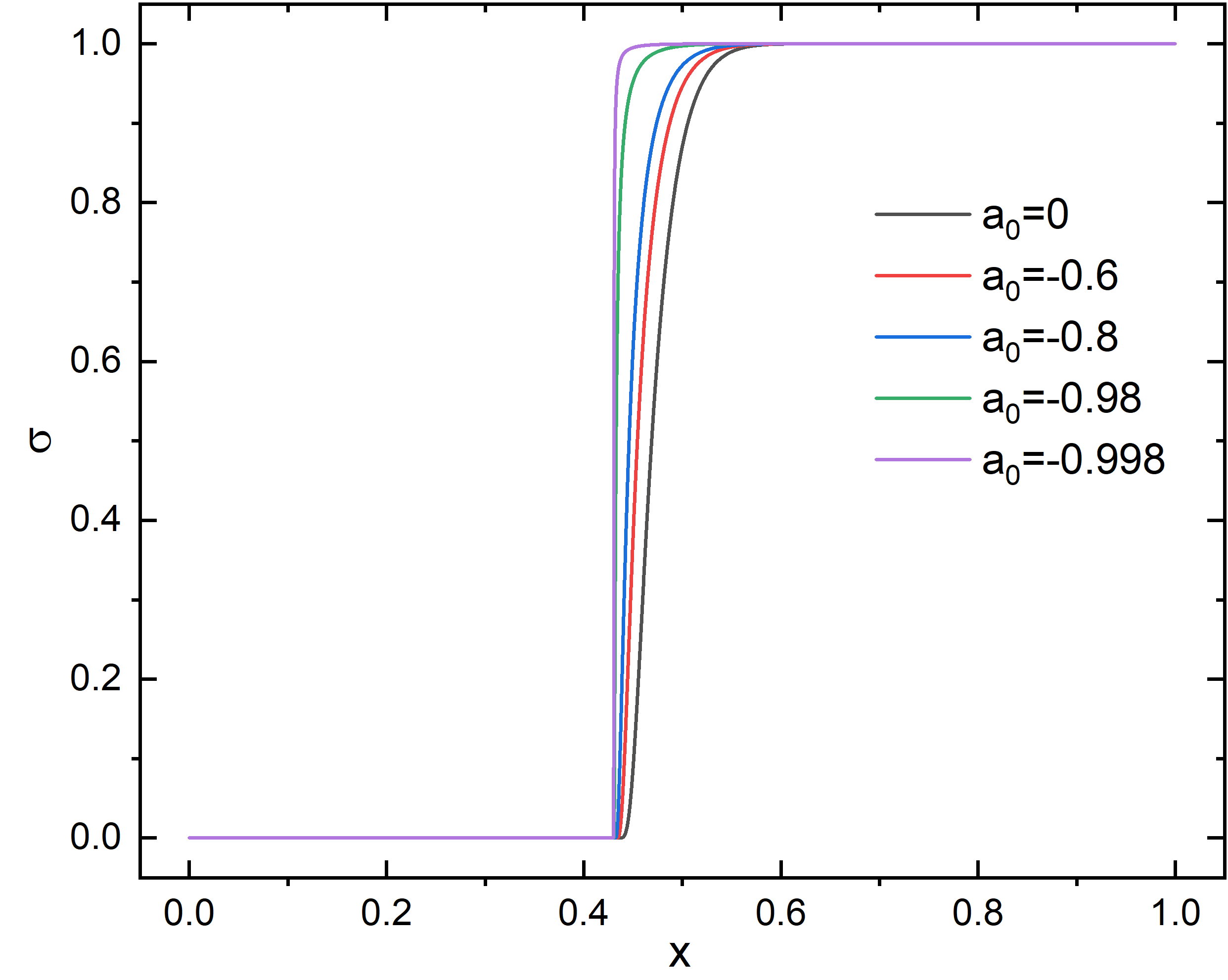}
  \end{center}
  \caption{The distributions of  $P_\perp$ and $\sigma(r)$ under different values of $a_0$ for the frozen state at $\rho_0 = 0.6839105$,  \;$n=0.5$, and  $h=0.5$.
  }\label{linjie}
\end{figure}

In Fig.~\ref{linjie}, we present the distributions for the frozen state at $\rho_0 = 0.6839105$ (nearly approaching the extremal black hole limit), focusing on the tangential pressure $P_\perp$ and the metric function $\sigma(r)$ under different values of $a_0$. The tangential pressure exhibits a sharp peak near the critical horizon, with the peak becoming narrower and more pronounced as $a_0$ approaches $-1$. Meanwhile, the $\sigma(r)$ function tends toward a step-like profile as $a_0$ decreases toward $-1$, reflecting a rapid transition in the redshift factor that effectively ``freezes'' time near the critical radius from the perspective of distant observers.

\subsection{Case II}

\begin{figure}[]
  \begin{center}
     \includegraphics[width=8.1cm]{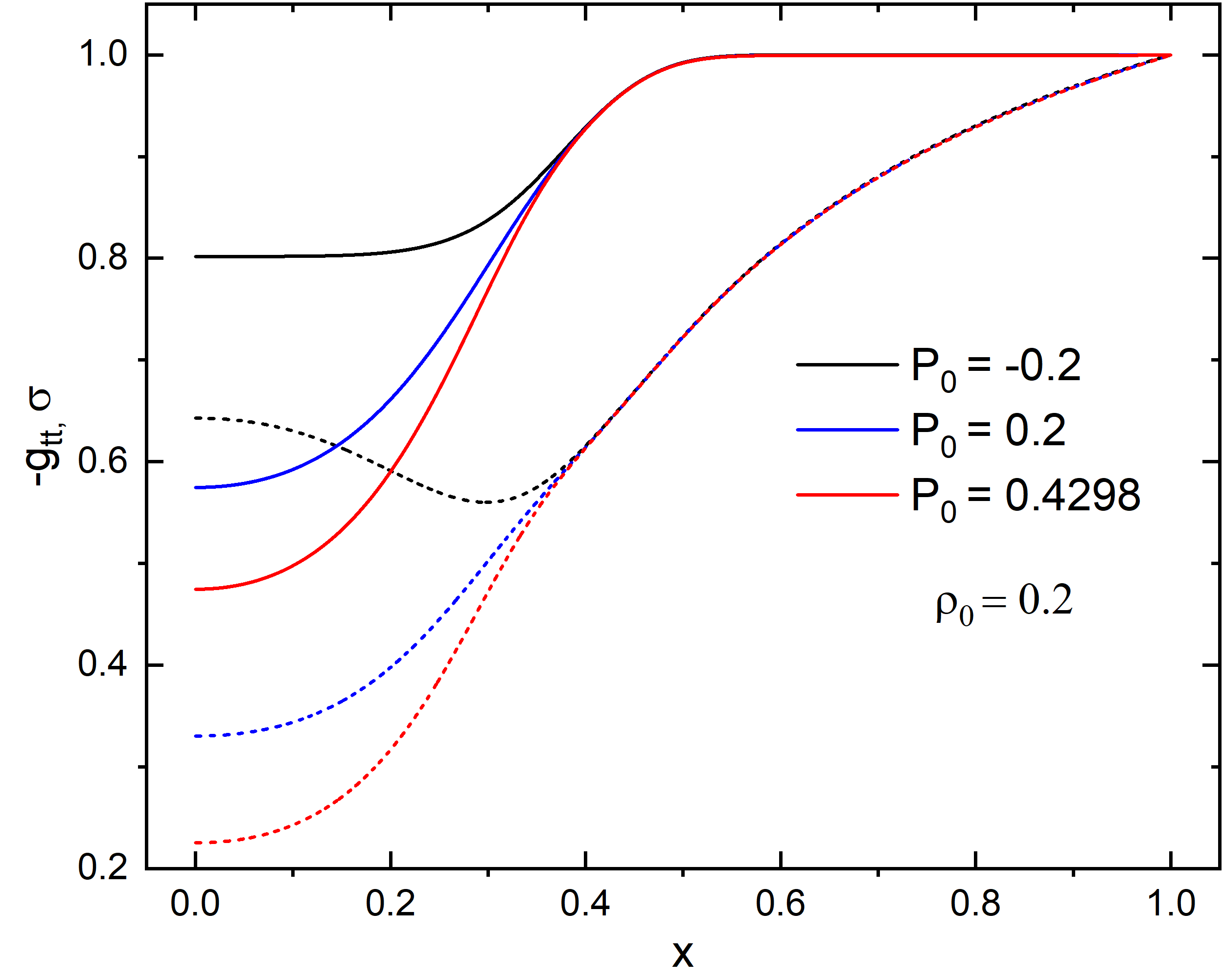}
   \includegraphics[width=8.1cm]{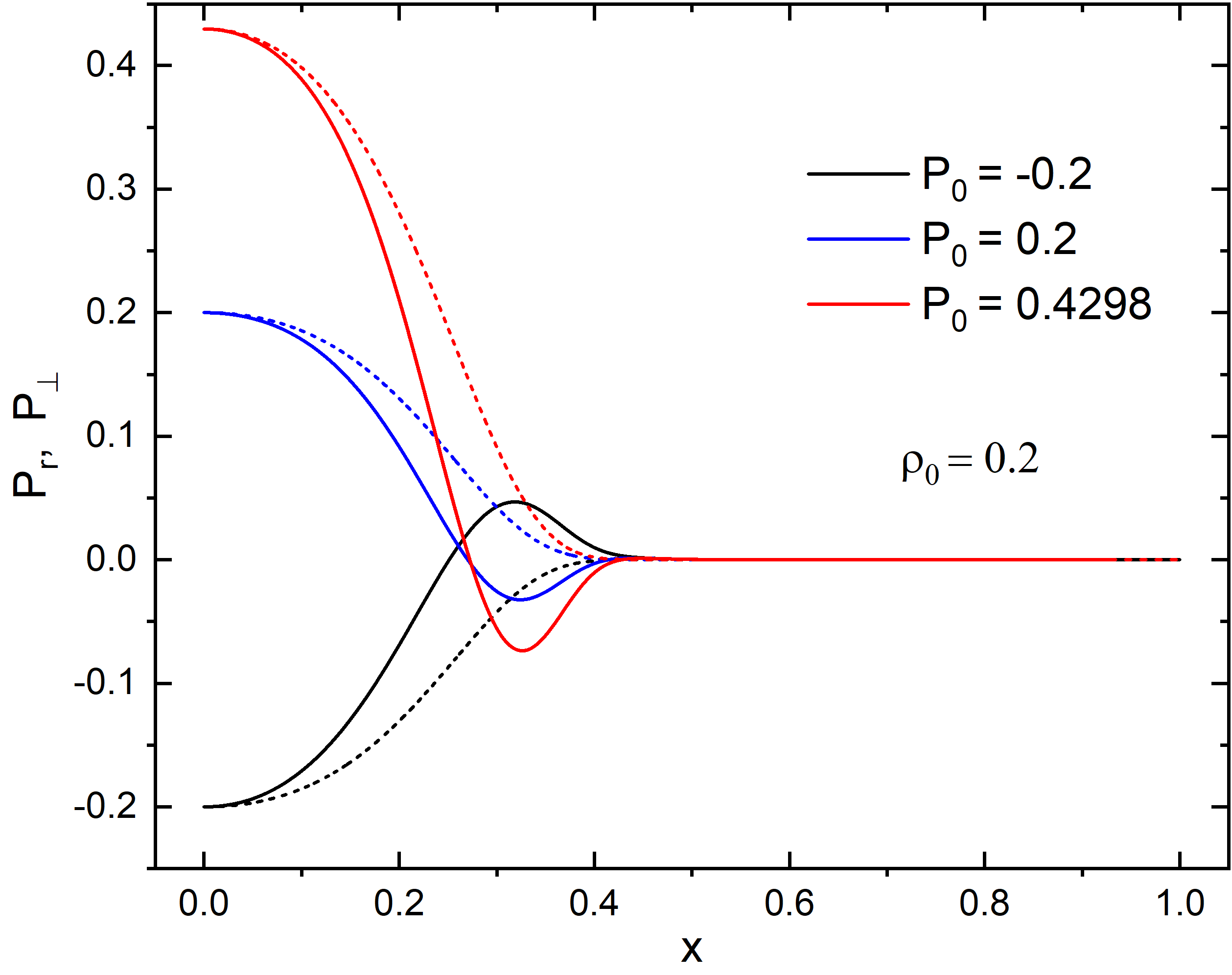}
    \includegraphics[width=8.4cm]{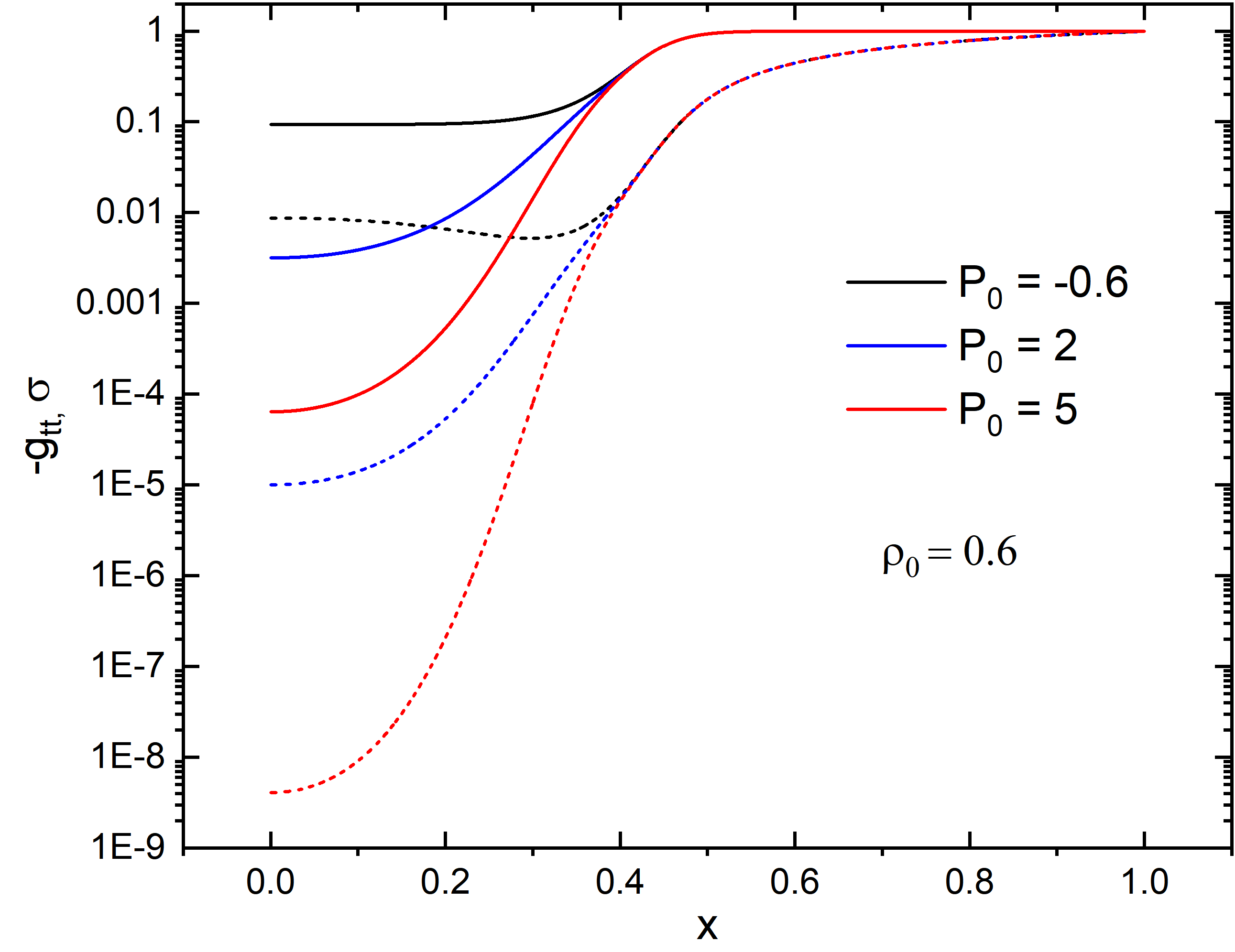}
   \includegraphics[width=7.9cm]{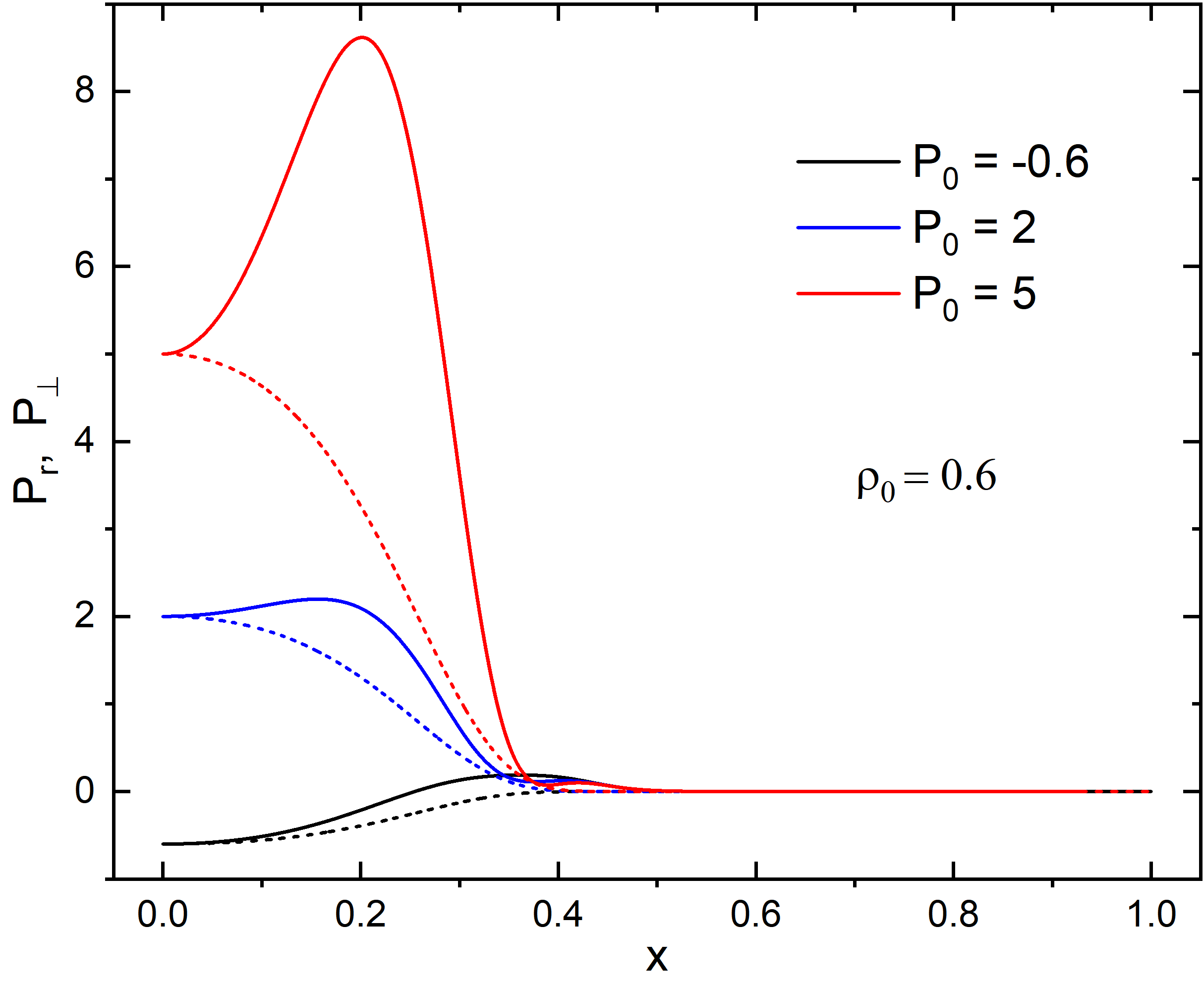}
  \end{center}
  \caption{Spatial profiles of the functions $-g_{tt}$ (dashed), $\sigma$ (solid), $P_r$ (dashed), and $P_\perp$ (solid) for two representative central densities, $\rho_0 = 0.2$ and $\rho_0 = 0.6$. The parameters are fixed as $n=1/2$, $h=1/2$, $a=2$, $m=4$, $b=1$, $k=1$, and $\gamma=6$.
  }\label{phaseca2}
\end{figure}

In this case, we adopt the radial pressure $P_r(r)$ as defined in Eq. (\ref{pr2}). In Fig.~\ref{phaseca2}, we illustrate the spatial profiles of the  functions $-g_{tt}$, $\sigma$, $P_r$, and $P_\perp$ for two representative central densities, $\rho_0 = 0.2$ and $\rho_0 = 0.6$, and examine their dependence on the parameter $P_0$.  For $\rho_0 = 0.2$,  there exists a maximum threshold for $P_0=0.4298$ beyond which the weak energy condition is violated.  In the case of $\rho_0 = 0.6$, a similar maximum threshold for $P_0$ exists, beyond which the weak energy condition fails.
Additionally, the minimum value of $-g_{tt}$ asymptotically approaches zero as $P_0$ increases.

\begin{figure}[]
  \begin{center}
  \includegraphics[width=8.1cm]{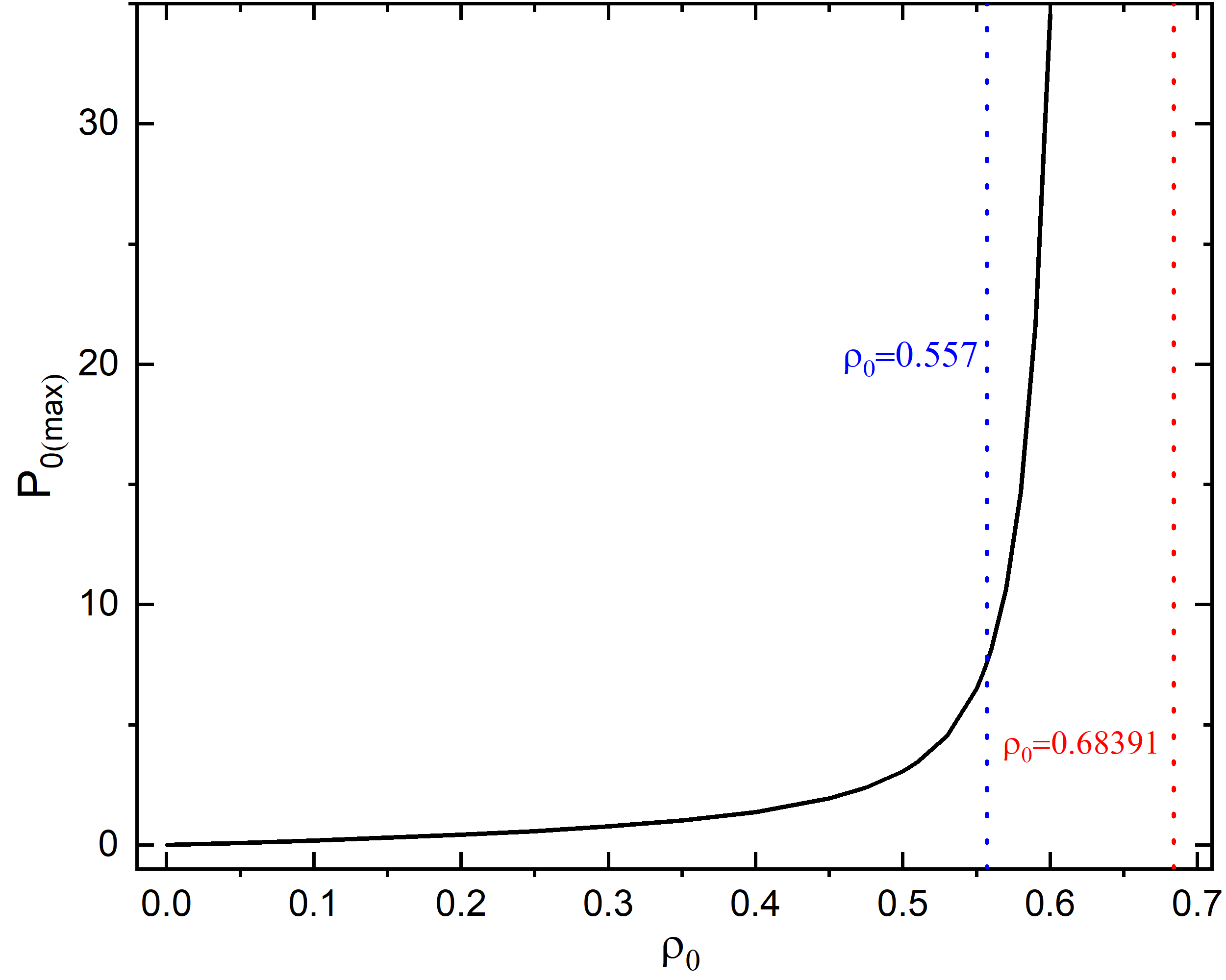}
  \end{center}
  \caption{The relationship between the maximum allowable $P_0$ that preserves the WEC and the central density $\rho_0$ with $n=1/2$, $h=1/2$, $ a=2$, $m=4$, $b=1$, $k=1$, $\gamma=6$.
  }\label{phmm2}
\end{figure}

Fig.~\ref{phmm2} shows the relationship between the maximum allowable $P_0$ that preserves the WEC and the central density $\rho_0$. It is evident that the upper bound of $P_0$ increases with $\rho_0$. As $\rho_0$ approaches the extremal black hole limit (represented by the red dashed line), the maximum $P_0$ consistent with the WEC tends toward infinity. Furthermore, the blue dashed line indicates a critical value $\rho_0 = 0.557$. At this point, under the maximum $P_0$, the minimum of $-g_{tt}$ drops below $10^{-10}$, a configuration we define as the ``frozen state." Our results suggest that a frozen state can only be realized when $\rho_0 \ge 0.557$; for densities below this threshold, no such state exists.

\section{Stability analysis of axial perturbations}
Following the Regge-Wheeler formalism, the axial-type perturbations can be reduced to a wave-like equation \cite{Chakraborty:2024gcr}:
\begin{equation}
    \frac{d^2 \Psi}{dr_*^2} + \left( \omega^2 - V(r) \right) \Psi = 0,
\end{equation}
where $r_*$ is the tortoise coordinate. The effective potential for the ``up'' and ``down'' perturbations are expressed as:
\begin{equation}
\begin{aligned}
V^{(\text{up})}(r) &= N \sigma^2\left( \frac{\ell(\ell+1)}{r^2} - \frac{6m(r)}{r^3} + 4\pi \left[ \rho(r) - 5P_r(r) + 4P(r) \right] \right), \\
V^{(\text{down})}(r) &= N \sigma^2 \left( \frac{\ell(\ell+1)}{r^2} - \frac{6m(r)}{r^3} + 4\pi \rho(r) - 4\pi P_r(r) \right).
\end{aligned}
\end{equation}

\begin{figure}[]
  \begin{center}
  \includegraphics[width=8.1cm]{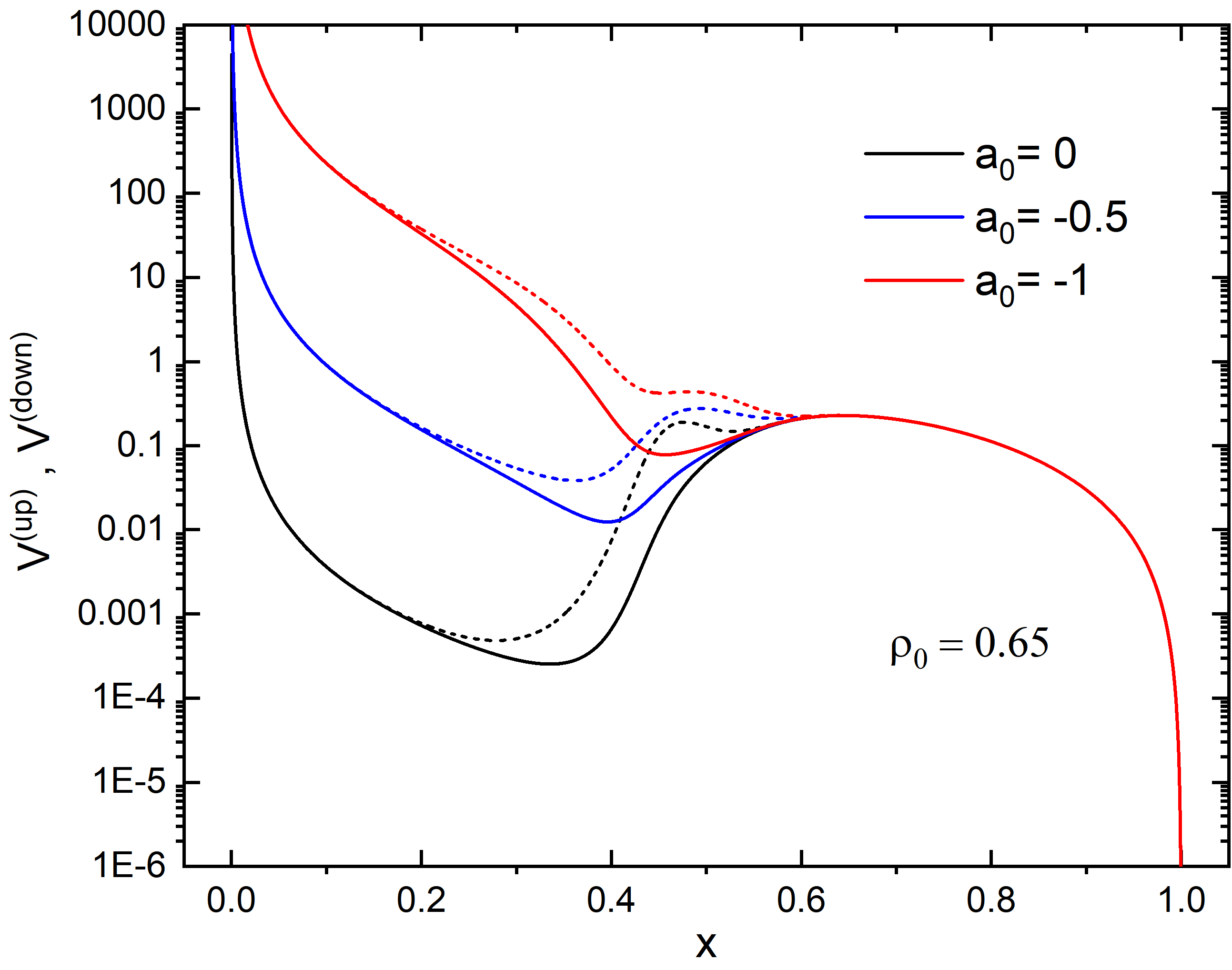}
   \includegraphics[width=8.1cm]{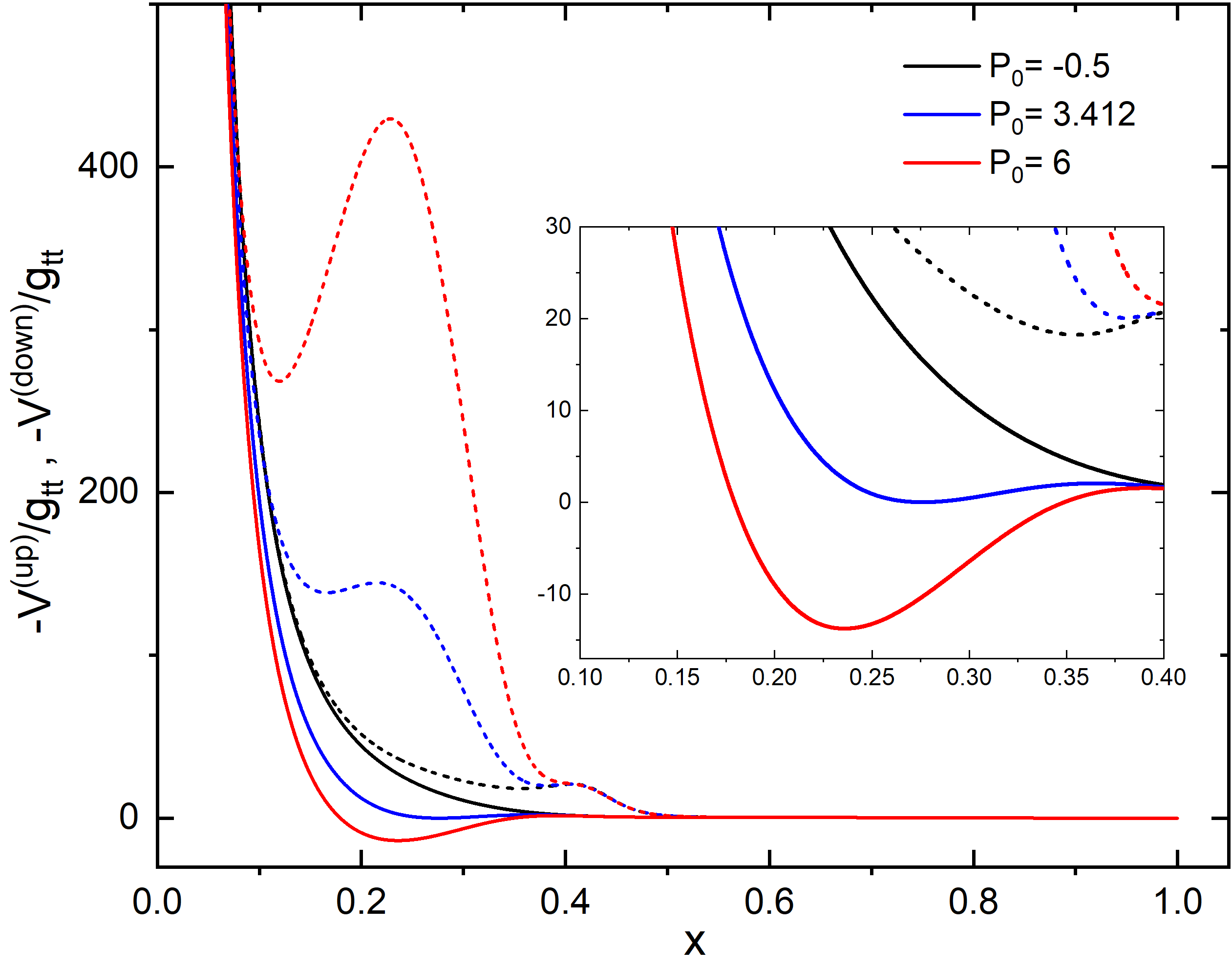}
  \end{center}
  \caption{The spatial evolution of the effective potentials $V^{(\text{up})}$ (dashed) and $V^{(\text{down})}$ (solid) with $\rho_0=0.65$, $n=1/2$, $h=1/2$, $a=2$, $m=4$, $b=1$, $k=1$, $\gamma=6$.}\label{stat12}
\end{figure}

In Fig.~\ref{stat12}, we illustrate the spatial evolution of the effective potentials, where the left and right panels correspond to Case I and Case II, respectively. The dashed lines represent $V^{(\text{up})}(r)$, while the solid lines denote $V^{(\text{down})}(r)$. For Case I, it is observed that both potential functions increase as the parameter $a_0$ decreases. Throughout the permissible range of $a_0$, the potentials remains strictly positive ($V(r) > 0$), thereby ensuring the linear stability of these compact star configurations against axial perturbations.
However, a distinct behavior is observed for Case II as the parameter $P_0$ increases. While $V^{(\text{up})}(r)$ remains positive, $V^{(\text{down})}(r)$ can develop negative regions. Specifically, as shown in the right panel of Fig.~\ref{stat12}, when $P_0 > 3.412$, $V^{(\text{down})}(r)$ drops below zero. The appearance of such negative potential wells may signal a transition in the stability properties of the system.

\section{Conclusions}
In this work, we have systematically investigated the gravitational configurations sourced by dark matter halos within a broader framework of anisotropic fluid dynamics. By relaxing the ad hoc constraint $P_r = -\rho$ commonly adopted in recent literature, we have obtained a more general class of self-consistent solutions to the Einstein field equations.
A important result of our analysis is the emergence of a transition between regular black holes and horizonless compact objects. We have shown that while the $P_r = -\rho$ case yields the well-known regular black hole geometries, more general anisotropic relations typically describe novel classes of compact stars. We identified a critical threshold in the parameter space, above which the system can reach a ``frozen state." In this regime, the asymptotic vanishing of the metric component $g_{tt}$ at finite radial coordinate signals the formation of a surface of infinite redshift—a phenomenon readily detectable by a distant static observer.

Furthermore, we performed a linear stability analysis against axial perturbations. We found that configurations in Case I remain stable throughout the permissible parameter range, characterized by strictly positive effective potentials. In contrast, for Case II, we identified that exceeding certain thresholds of the central pressure $P_0$ leads to negative potential wells, signaling potential instabilities.

In conclusion, our study provides a self-consistent theoretical framework that bridges small-scale black hole physics with large-scale galactic DM distributions. By circumventing the inconsistencies identified in prior models, this generalization offers a more robust phenomenological route for exploring the nature of dark matter in strong-gravity regimes and the search for horizonless compact mimickers in the universe.

\section{Acknowledgment}
This work is supported by the National Natural Science Foundation of China (Grants No.~12275110 and No.~12247101) and National Key Research and Development Program of China (Grant No. 2020YFC2201503).


\begin{thebibliography}{99}
\bibitem{Penrose:1965}
R.~Penrose,
``Gravitational Collapse and Space-Time Singularities,''
\href{https://doi.org/10.1103/PhysRevLett.14.57}{Phys. Rev. Lett.} \textbf{14}, 57 (1965).

\bibitem{Hawking:1970}
S.~W.~Hawking and R.~Penrose,
``The Singularities of Gravitational Collapse and Cosmology,''
\href{https://doi.org/10.1098/rspa.1970.0021}{Proc. R. Soc. London A} \textbf{314}, 529 (1970).

\bibitem{Bardeen:1968}
J.~M.~Bardeen,
in \textit{Proceedings of GR5, Tbilisi, USSR} (1968)

\bibitem{Hayward:1994}
S.~A.~Hayward,
``General laws of black-hole dynamics,''
\href{https://doi.org/10.1103/PhysRevD.49.6467}{Phys. Rev. D} \textbf{49}, 6467 (1994).

\bibitem{Ayon-Beato:1998hmi}
E.~Ayon-Beato and A.~Garcia,
``Regular black hole in general relativity coupled to nonlinear electrodynamics,''
Phys. Rev. Lett. \textbf{80}, 5056-5059 (1998)
doi:10.1103/PhysRevLett.80.5056
[arXiv:gr-qc/9911046 [gr-qc]].


\bibitem{Ayon-Beato:2000mjt}
E.~Ayon-Beato and A.~Garcia,
``The Bardeen model as a nonlinear magnetic monopole,''
Phys. Lett. B \textbf{493}, 149-152 (2000)
doi:10.1016/S0370-2693(00)01125-4
[arXiv:gr-qc/0009077 [gr-qc]].

\bibitem{Bueno:2024dgm}
P.~Bueno, P.~A.~Cano and R.~A.~Hennigar,
``Regular black holes from pure gravity,''
Phys. Lett. B \textbf{861}, 139260 (2025)
doi:10.1016/j.physletb.2025.139260
[arXiv:2403.04827 [gr-qc]].

\bibitem{Bueno:2025zaj}
P.~Bueno, P.~A.~Cano, R.~A.~Hennigar and {\'A}.~J.~Murcia,
``Regular black hole formation in four-dimensional nonpolynomial gravities,''
Phys. Rev. D \textbf{113}, no.2, 024019 (2026)
doi:10.1103/8f3j-zcxh
[arXiv:2509.19016 [gr-qc]].


\bibitem{Konoplya:2025}
R.~A.~Konoplya and A.~Zhidenko,
``Dark matter halo as a source of regular black-hole geometries,''
\href{https://arxiv.org/abs/2511.03066}{arXiv:2511.03066v2 [gr-qc]} (2025).



\bibitem{Bolokhov:2025qnm}
S.~V.~Bolokhov,
``Quasinormal ringing of a regular black hole sourced by the Dehnen-type distribution of matter,''
\href{https://arxiv.org/abs/2511.12859}{arXiv:2511.12859v1 [gr-qc]} (2025).

\bibitem{Saka:2025}
E.~U.~Saka,
``Regular black hole sourced by the Dehnen-type distribution of matter: The sound of the event horizon,''
\href{https://arxiv.org/abs/2512.08904}{arXiv:2512.08904v1 [gr-qc]} (2025).

\bibitem{Malik:2025}
Z.~Malik,
``Long-lived quasinormal modes and grey-body factors of supermassive black holes with a dark matter halo,''
\href{https://arxiv.org/abs/2511.12335}{arXiv:2511.12335v1 [gr-qc]} (2025).



\bibitem{Lutfuoglu:2025grav}
B.~C.~L{\"u}tf{\"u}o\u{g}lu \textit{et al.},
``Gravitational Spectra and Wave Propagation in Regular Black Holes Supported by a Dehnen Halo,''
\href{https://arxiv.org/abs/2511.22366}{arXiv:2511.22366v1 [gr-qc]} (2025).


\bibitem{Bolokhov:2025crit}
S.~V.~Bolokhov,
``Revisiting black holes in dark-matter halos: on consistent solutions to the Einstein equations,''
\href{https://arxiv.org/abs/2512.06930}{arXiv:2512.06930v1 [gr-qc]} (2025).





\bibitem{Wang:2023tdz}
X.~E.~Wang,
``From Bardeen-boson stars to black holes without event horizon,''
[arXiv:2305.19057 [gr-qc]].

\bibitem{Yue:2023sep}
Y.~Yue and Y.~Q.~Wang,
``Frozen Hayward-boson stars,''
JCAP \textbf{05}, 066 (2025)
[arXiv:2312.07224 [gr-qc]].

\bibitem{Brihaye:2025dlq}
Y.~Brihaye and B.~Hartmann,
``Frozen states of charged boson stars,''
[arXiv:2507.08946 [gr-qc]].

\bibitem{Chicaiza-Medina:2025wul}
S.~S.~Chicaiza-Medina and J.~C.~Degollado,
``Hayward Boson Stars,''
[arXiv:2508.11906 [gr-qc]].


\bibitem{Tan:2025jcg}
C.~Tan and Y.~Q.~Wang,
``Frozen Neutron Stars,''
[arXiv:2509.09338 [gr-qc]].

\bibitem{Einasto1965}
J. Einasto, Trudy Astrofizicheskogo Instituta Alma-Ata \textbf{5}, 87 (1965), in Russian.

\bibitem{EinastoHaud1989}
J. Einasto and U. Haud, Astronomy and Astrophysics \textbf{223}, 89 (1989).

\bibitem{Retana2012}
E. Retana-Montenegro, G. Gentile, M. Baes, F. Frutos-Alfaro, and E. Van Hese, \textit{Astron. Astrophys.} \textbf{540}, A70 (2012), arXiv:1202.5242 [astro-ph.CO].

\bibitem{Chakraborty:2024gcr}
S.~Chakraborty, G.~Comp{\`e}re and L.~Machet,
``Tidal Love numbers and quasinormal modes of the Schwarzschild-Hernquist black hole,''
Phys. Rev. D \textbf{112}, no.2, 024015 (2025)
doi:10.1103/4p2c-rwdh
[arXiv:2412.14831 [gr-qc]].
\end{thebibliography}
\end{document}